\documentclass[prb,aps,twocolumn,superscriptaddress]{revtex4-1}
\usepackage{graphicx,amssymb,amsmath,color,psfrag}
\usepackage{amsthm}
\usepackage{amsfonts}
\usepackage{algorithmic}
\usepackage{enumerate}
\usepackage{latexsym}

\def\avg#1{\left\langle#1\right\rangle}

\def\be{\begin{equation}}       \def\ee{\end{equation}}
\def\bea{\begin{eqnarray}}      \def\eea{\end{eqnarray}}
\def\ba{\begin{array} }
\def\ea{\end{array} }
\def\bnum{\begin{enumerate} }
\def\enum{\end{enumerate}}

\def\=>{\Rightarrow}
\def\>{\rightarrow}

\def\eye2{Fathbb{I}}

\renewcommand{\v}[1]{{\bf #1}}

\renewcommand{\>}{\rangle}

\begin{document}

\title{Strain induced edge magnetism at the zigzag edge of a graphene quantum dot}

\author{Shuai Cheng}
\affiliation{Department of Physics, Beijing Normal University,
Beijing 100875, China\\}

\author{Jinming Yu}
\affiliation{Department of Physics, Beijing Normal University,
Beijing 100875, China\\}

\author{Tianxing Ma}
\email{txma@bnu.edu.cn}
\affiliation{Department of Physics, Beijing Normal University,
Beijing 100875, China\\}
\affiliation{Beijing Computational Science Research Center,
Beijing 100084, China}

 \author{N. M. R. Peres}
\affiliation{Centro de F\'{\i}sica and Departamento de F\'{\i}sica, Universidade
do Minho, Campus de Gualtar, Braga 4710-057, Portugal}


\begin{abstract}
We study the temperature dependent magnetic susceptibility of a strained graphene quantum dot
using the determinant quantum Monte Carlo method.
Within the Hubbard model on a honeycomb lattice, our unbiased numerical results show that a relative
small interaction $U$ may lead to a edge ferromagnetic-like behavior in the strained graphene quantum dot.
Around half filling, the ferromagnetic fluctuations at the zigzag edge are strengthened both
by the on-site Coulomb interaction and the strain, especially in low temperature region.
\end{abstract}
\maketitle
 
\section{Introduction}
\begin{figure}[tbp]
\includegraphics[scale=0.4]{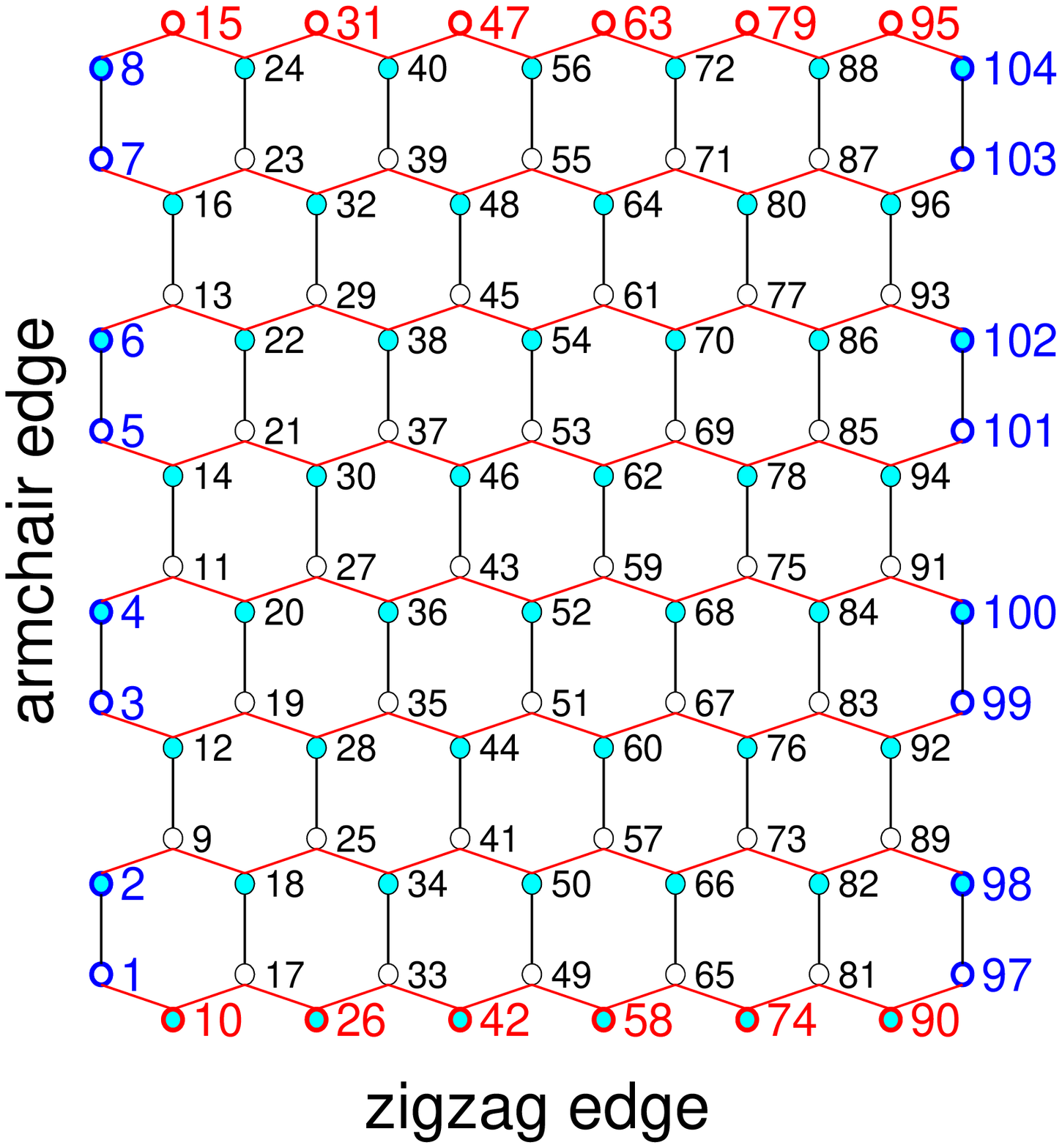}
\caption{(Color online) A sketch for a graphene quantum dot with 104 sites
where white and cyan circles indicates A and B sublattice respectively. The sites at zigzag edge are marked by red color numbers and the sites at armchair edge are marked by blue color numbers.
We consider the strain along the zigzag-direction.
The dark line indicates $t_1=t$, red lines indicates $t_2=t_3=t-\Delta t$.
 Here $t$ represents the nearest hoping term and $\Delta t$ represents the effect of strain. }
\label{Fig:structure}
\end{figure}

Graphene-based systems have been the subject of a considerable body of
research\cite{Novoselov2004,Novoselov2005,Neto2009,Wassmann2008,Yang2008}
due to their potential application in nano-electronic
devices
\cite{Yan2010,Yazyev2008,MaAPLs,Rossier2007,Bhowmick2008,Jiang2008,Yazyev2010,Peres2005,Peres2010,Nair2012,Sharma2013,Roy2014}.
A perfect graphene sheet consists of a single layer of carbon atoms arranged in a
honeycomb crystal lattice  as depicted in Fig.
\ref{Fig:structure}.
Since its
discovery, graphene research expanded quickly, and
graphene-based systems with different edge topology have been synthesised.
It has been suggested that the electronic properties of graphene quantum dots
with different edges may find interesting applications in
nano-electronic
devices, where their edge structure --zigzag, armchair, or something in between-- will provide
different routes to specific applications.
The graphene-based quantum dot depicted in Fig. \ref{Fig:structure} shows two
different types of edges --zig-zag and armchair. For a graphene nanoribbon
one can assume it to be infinite in one direction but finite in the perpendicular one.
In this way one can produce  graphene nanoribbons  with either zigzag or armchair
terminations\cite{Neto2009}. For a quantum dot, and excluding very specific cases, one always
have, at least, the two types of terminations present. That is the case we consider in this
paper.

The possibility magnetism in graphene-based materials is an important problem and
may open new avenues toward the  development of spintronics
\cite{Rossier2007,Bhowmick2008,Jiang2008,Yazyev2010,Peres2005,Nair2012,Sharma2013,Roy2014,Golor2014}.
In general,
 spintronics\cite{spintronics1} requires a
 semiconductor material with some type of magnetic property at (or above) room temperature\cite{FE}.
In perfect graphene, it was suggested that antiferromagnetic correlations dominate
 around half filling, and  ferromagnetic fluctuations
may dominate in a rather high filling (doped) region around the Van Hove singularity
in the density of state\cite{Ma2010}. Unfortunately this level of doping is still
far from the current experimental
ability to dope the material\cite{Filling,VHS}.
The possible ferromagnetic order that was proposed to exist in graphene-based materials with defects,
such as vacancies, topological
defects, and hydrogen chemiadsorption, are all waiting for experimental confirmation
\cite{Yazyev2008B,Ugeda2010,Giesbers2013}.
 
Graphene nanoribbons' magnetism has also attracted considerable attention, since it holds
promises of many applications in the design of nanoscale magnetic
and spintronics devices. It has been shown that the zigzag graphene nanoribbons
exhibit  ferromagnetic correlations along the edge at half filling\cite{Feldner2011}, and
that armchair
graphene nanoribbons have ferromagnetic fluctuations in the
doped region around the nearly flat band\cite{Ma2012}.

The shape and symmetry
of the dots  play an
important role in the energy level statistics and in the spatial
charge density\cite{Wurm2009,Akola2008}.
These properties, spur the interest in magnetism in graphene quantum dots. The tight-binding description of graphene quantum dots reveals that the structure of the
edge-state spectrum and the magnetic response of the dots is
strongly dependent on the geometric shape of the cluster. Indeed, it exists the possibility of
crossover  between paramagnetic and
diamagnetic responses of the system as a function of its shape, size,
and temperature\cite{Espinosa2013}. 
The possibility of ground state magnetization in strained graphene quantum dots
was suggested by mean-filed calculations\cite{Peres2009}, which revealed that
magnetism can be enhanced by as much as 100\% for strain values on the order of 20\%.

The mean field results show that the critical Hubbard
interaction $U_c$ for bulk graphene (unstrained) is about $2.23 t$, where $t$
is the nearest hoping term of the honeycomb lattice.
This value of $U_c$ put the system into a  moderate
correlated regime, as $U_c$ is near to the half bandwidth $w$, where $w$
is about $6t$\cite{Ma2010}. For such a $U_c$ value, the mean field method may lead to
spurious results because the system is very sensitive to the approximation used.
The temperature dependent magnetic susceptibility plays a
key role in understanding the behavior of magnetism and is used in this paper as
a probe to magnetic correlations in graphene quantum dots.

In this paper, using an unbiased numerical method, we study the temperature
dependent magnetic susceptibility in
a strained graphene quantum dot.

\section{Model and methods}
Strain is an active topic of experimental research both in semiconductors in general
and in graphene-based materials.
Some degree of strain can be
induced either by deposition of oxide capping layers  or by mechanical methods\cite{Peres2009}.
In the present work, we concentrate on the half-filled and low doping regimes of a
graphene quantum dot, a doping level that can be easily realized in experiments\cite{VHS}.
Our numerical results reveal a high-temperature ferromagnetic-like behavior at
the edges of a strained graphene quantum dot, for reasonable interaction electron-electron interaction
values. Such ferromagnetic correlations are enhanced by increasing both the
strain and the interaction strength.

Fig. \ref{Fig:structure} depicts the system under study, which
is an honeycomb  lattice with 8$\times 13$ sites.  We can change the size of lattice by
changing the length along each edge.
The sites at armchair edges have been marked with blue numbers and the sites at the zig-zag ones
have been marked with red numbers.

The Hamiltonian for  a stained graphene quantum dot can be expressed as
\begin{eqnarray}
H=&&\sum_{\mathbf{i\eta}\sigma}t_{\eta} a_{\mathbf{i}\sigma}^{\dag}b_{%
\mathbf{i+\eta}\sigma}+ h.c. + U\sum_{\mathbf{i}}(n_{\mathbf{ai}\uparrow}n_{\mathbf{ai}\downarrow}+n_{\mathbf{bi}\uparrow}n_{\mathbf{bi}\downarrow}) \notag \\
&&+\mu\sum_{\mathbf{i}\sigma}(n_{\mathbf{ai}\sigma }+n_{\mathbf{bi}\sigma})
\end{eqnarray}
Here, $a_{i\sigma}$ ($a_{i\sigma}^{\dag}$) annihilates (creates)
electrons at site $\mathbf{R}_i$ with spin $\sigma$ ($\sigma$=$\uparrow,\downarrow$)
on sublattice A, as well as $b_{i\sigma}$ ($b_{i\sigma}^{\dag}$) acting
on electrons of sublattice B, $n_{ai\sigma}=a_{i\sigma}^{\dagger}a_{i\sigma}$
and $n_{bi\sigma}=b_{i\sigma}^{\dagger}b_{i\sigma}$.
$U$ is the on-site Hubbard interaction and $\mu$ is the chemical potential.
On such honeycomb lattice, $t_{\eta}$ denotes the nearest neighbor hoping integral.
We consider that stress is applied  along the zigzag direction. The applied stress
modifies the interatomic distances, which in
turn implies a change in the electronic-hopping parameters $t_{\eta}$.
As a consequence of these changes
the band structure of the material is modified.
The quantitative change in the hoppings
upon stress was studied using ab initio methods, and we illustrate that in Fig. \ref{Fig:structure}.
The lines in dark indicating hoping terms with $t_{1}=t$ along the direction of stress,
which do not change in value.
The lines in red change their values as $t_{2,3}=t-\Delta t$,
according to the strength of stress parametrized by $\Delta t$.

The nearest-neighbor hopping energy $t$ reported in the
literature\cite{Neto2009} ranges from 2.5 to 2.8 eV, 
and the value of the on-site repulsion $U$ can be taken from its
estimation in polyacetylene\cite{Parr1950,Herbut2006,Neto2009}
--U$\cong$6.0-17 eV, which clearly spans a large range of values.

In principle it is questionable to apply for correlated electrons in graphene
 the simplest version of
the Hubbard model with values of $U$ valid for polyacetylene.
However, the Peierls-Feynman-Bogoliubov
variational principle shows that a generalized Hubbard model with non-local Coulomb
interactions is mapped onto an effective Hubbard model with on-site effective
interaction $U$ only, which is about $1.6|t|$\cite{Schuler2013}. Following the latter reference
we
study the the model Hamitonian in the range of $U/|t|=1-3$. Although the
 value of $U/|t|=3$ is larger than $1.6|t|$, our aim is to explore the importance of
interactions on the magnetism of quantum dot under study. 

For such ranges of $U$ and $t$, the
 the determinant quantum Monte Carlo (DQMC) simulation is a reliable tool for
investigating the nature of magnetic correlations in the presence of  moderate Coulomb
 interactions. This is specially true in what concerns  changes of
the band structure with respect to
modifications of transverse width and to the edge topology.

In DQMC, the basic strategy  is to express the partition function as a high-dimensional integral over a set of
random auxiliary fields. Then the integral is accomplished by
Monte Carlo techniques. In present simulations,
8000 sweeps were used to equilibrate the system, and an
additional 30000 sweeps were  made, each of which generated a
measurement. These measurements were split into ten bins which provide the
basis of coarse-grain averages, and errors were estimated based on standard
deviations from the average. For more technique details, we refer to
Refs.~\cite{Blankenbecler1981,dqmcma}.

\section{Results}
To explore the behavior of magnetism in the graphene quantum dot,
we calculate the uniform magnetic susceptibility $\chi$ for the bulk,
the magnetic susceptibility $\chi_a$ at the armchair edge and the
magnetic susceptibility $\chi_z$ at the zigzag edge. Here
\begin{eqnarray}
\chi= \int_{0}^{\beta}d\tau \sum_{d,d'=a,b} \sum_{i,j}
\langle\textrm{m}_{i_{d}}(\tau) \cdot
\textrm{m}_{j_{d'}}(0)\rangle
\end{eqnarray}
where $m_{i_{a}}(\tau)$=$e^{H\tau} m_{i_{a}}(0) e^{-H\tau}$ with
$m_{i_{a}}$=$a^{\dag}_{i\uparrow}a_{i\uparrow}-a^{\dag}_{i\downarrow}a_{i\downarrow}$
and
$m_{i_{b}}$=$b^{\dag}_{i\uparrow}b_{i\uparrow}-b^{\dag}_{i\downarrow}b_{i\downarrow}$.
We measure $\chi$ in unit of $\mid t\mid^{-1}$.
The $\chi$ of the bulk is calculated by summing over all the sites.
The $\chi_a$ at the armchair edge is calculated by summing over
the sites marked with red-color numbers in Fig. \ref{Fig:structure},
and the $\chi_z$ at the zigzag edge is calculated by summing over the sites marked with
 blue-color numbers in the same figure.
An average for $\chi$, $\chi_a$, and $\chi_z$ is
 made corresponding to the respective total number of sites.

\begin{figure}[tbp]
\includegraphics[scale=0.45]{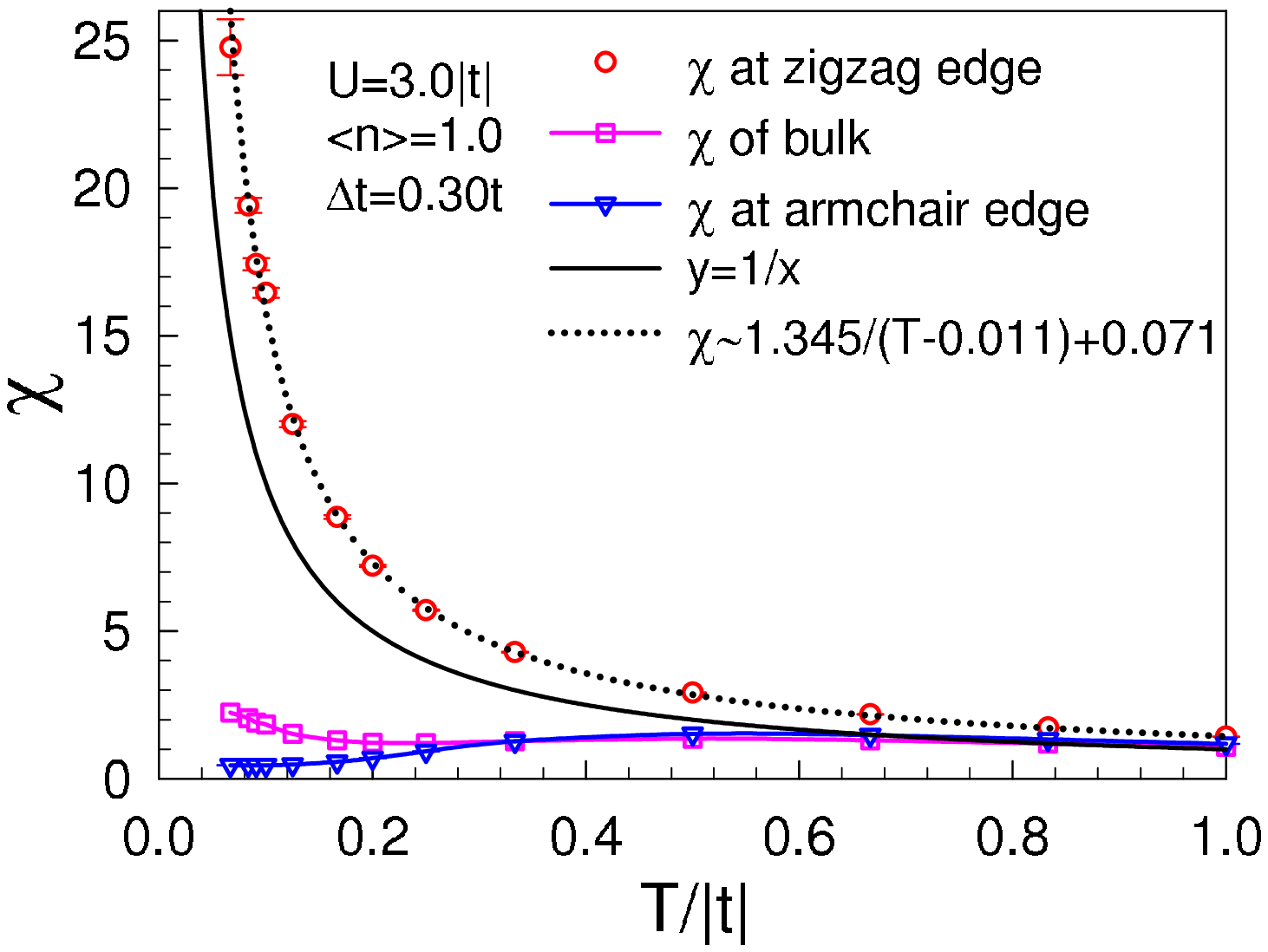}
\caption{(Color online) The $\chi_z$ (red circles), $\chi$ (pink line with square ), and
$\chi_a$ (blue lines with down triangle) as a function of temperature at $U=3.0$$\mid$$t$$\mid$, $\avg{n}=1.0$, and $\Delta t=0.30t$  of a lattice with $104$ sites.
}\label{L8n1U3st0.50}
\end{figure}

Firstly we present the temperature dependent $\chi$, $\chi_a$, and $\chi_z$ for $U=3.0\mid t \mid$,
$\avg{n}=1.0$, and $\Delta t=0.30t$ in Fig. \ref{L8n1U3st0.50}.
To qualitatively estimate the behavior of the temperature dependence of
the  magnetic susceptibility, we plot the function $y=1/x$, since
 the Curie-Weiss law --$\chi=C/(T-T_c)$-- describes the magnetic susceptibility $\chi$ for
a ferromagnetic material in the 
temperature region above the Curie temperature $T_c$.

We note that the $\chi_z$ (red circles) increases as the temperature decreases, which shows a
ferromagnetic-like behavior.
Interesting enough, the $\chi_a$ decreases as the temperature decreases.
As $\chi_z$ is much larger than the $\chi_a$,  the bulk
uniform magnetic susceptibility $\chi$ also increases as the temperature decreases,
especially in low temperature region.  Within our numerical results,
we fit the DQMC data with a formula of
\begin{eqnarray}
\label{Fit1}
\chi_z(T)=a/(T-T_c)+b
\end{eqnarray}
,as that shown (dashed lines)
in Fig. \ref{L8n1U3st0.50} which allows to estimate the  transition temperature $T_c$.
The fitting agrees with the DQMC data quite well.
From this fitting, one may estimate a $T_c$ of about $\sim 0.011t$, which is roughly $\sim 320$ K. For lower temperatures, one can notice significant error bars on the susceptibility, related to the Monte-Carlo sampling.
From Eq. \ref{Fit1}, we have
\begin{eqnarray}
\label{Fit2}
T_c=a/[\chi_z(T)-b]+T\,.
\end{eqnarray}
To estimate the error bar of the obtained $T_c$, we use the standard rule for
estimating errors of indirect
measurement by deriving the partial derivative of the right part of Eq.\ref{Fit2},
thus obtaining
 \begin{eqnarray}
\label{Fit3}
\delta T_c=a\delta \chi_z(T)/\chi^2_z(T)\,.
\end{eqnarray}
We use the susceptibility at the lowest temperature, $T_{lowest}$,
to estimate the error.
We can then
estimate $\delta T_c=a\delta \chi_z(T_{lowest})/\chi^2_z(T_{lowest})\simeq 0.002$$\mid$$t$$\mid$,
which indicates that the value of $T_c$ should be statistically distinguishable from zero. 

\begin{figure}[tbp]
\includegraphics[scale=0.45]{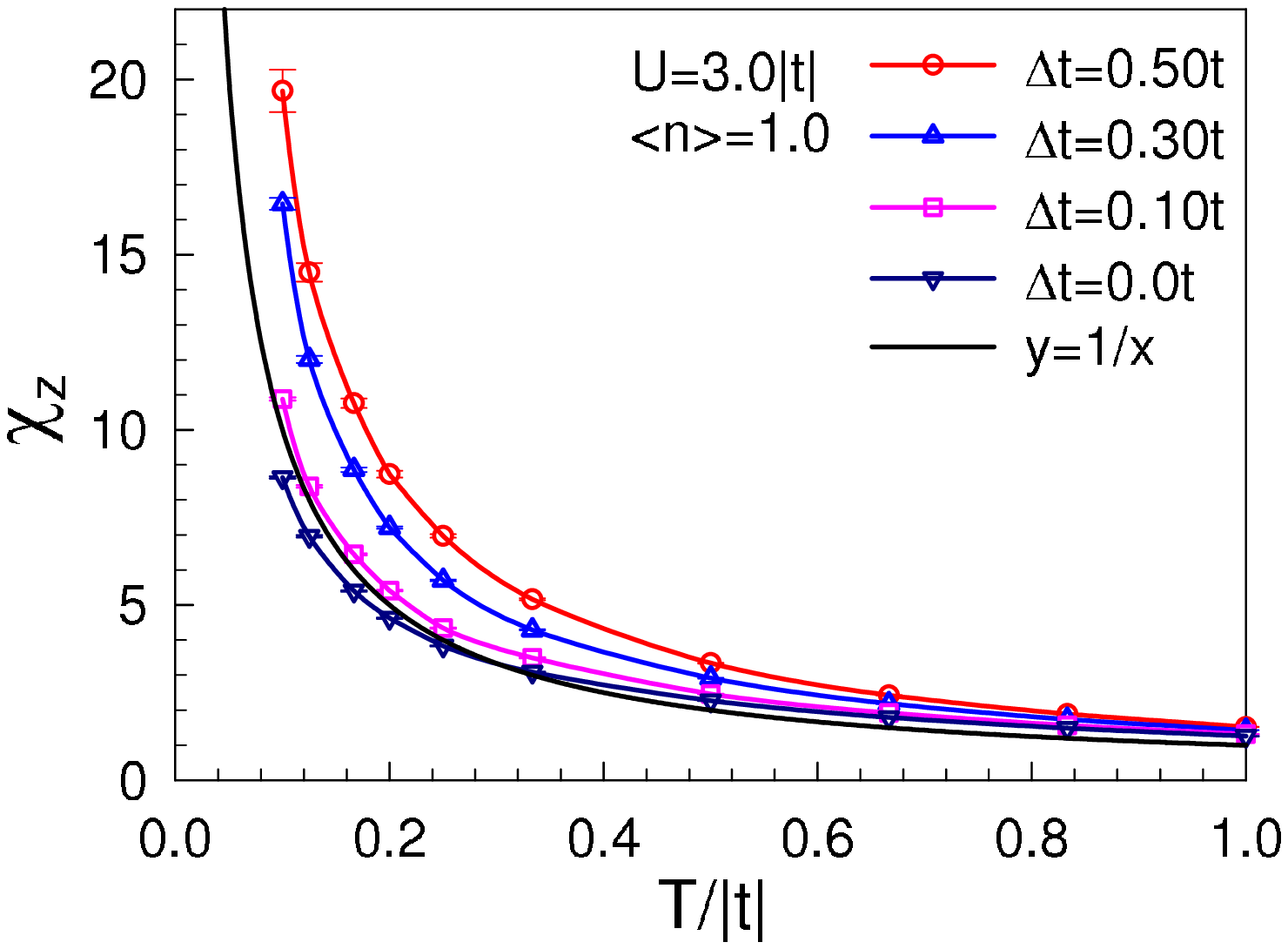}
\caption{(Color online) The $\chi_z$ at $U=3.0$$\mid$$t$$\mid$ and $\avg{n}=1.0$ with different strain.}
\label{strain}
\end{figure}

The difference between the temperature dependence of
$\chi_z$ and $\chi_a$ is due to the edge geometry. For an
half-filled Hubbard model on a perfect honeycomb lattice, the system
shows antiferromagnetic correlations. As the structure of the
honeycomb lattice can be described by two inter-penetrating
sub-lattices, the spin correlation between the nearest neighbour
sites is negative (due to antiferromagnetic correlations), and the spin correlation
between the next nearest neighbour sites belonging to the same
sub-lattice, has to be positive. In the graphene dot under study,
the sites along the armchair edge belong to different sub-lattices,
while the sites along the zig-zag edge belong to the same
sublattice. Thus, the magnetic susceptibility at the armchair edge
is antiferromagnetic-like while the magnetic susceptibility at the
zigzag edge is ferromagnetic-like. As noted already, the
susceptibility at the armchair edge is a non-monotonic function of
temperature. This may be caused by the competition between the
enhanced spin polarization with lowering temperature and unbalanced
distribution of electron with different spins at armchair and zigzag
edges. 

\begin{figure}[tbp]
\includegraphics[scale=0.45]{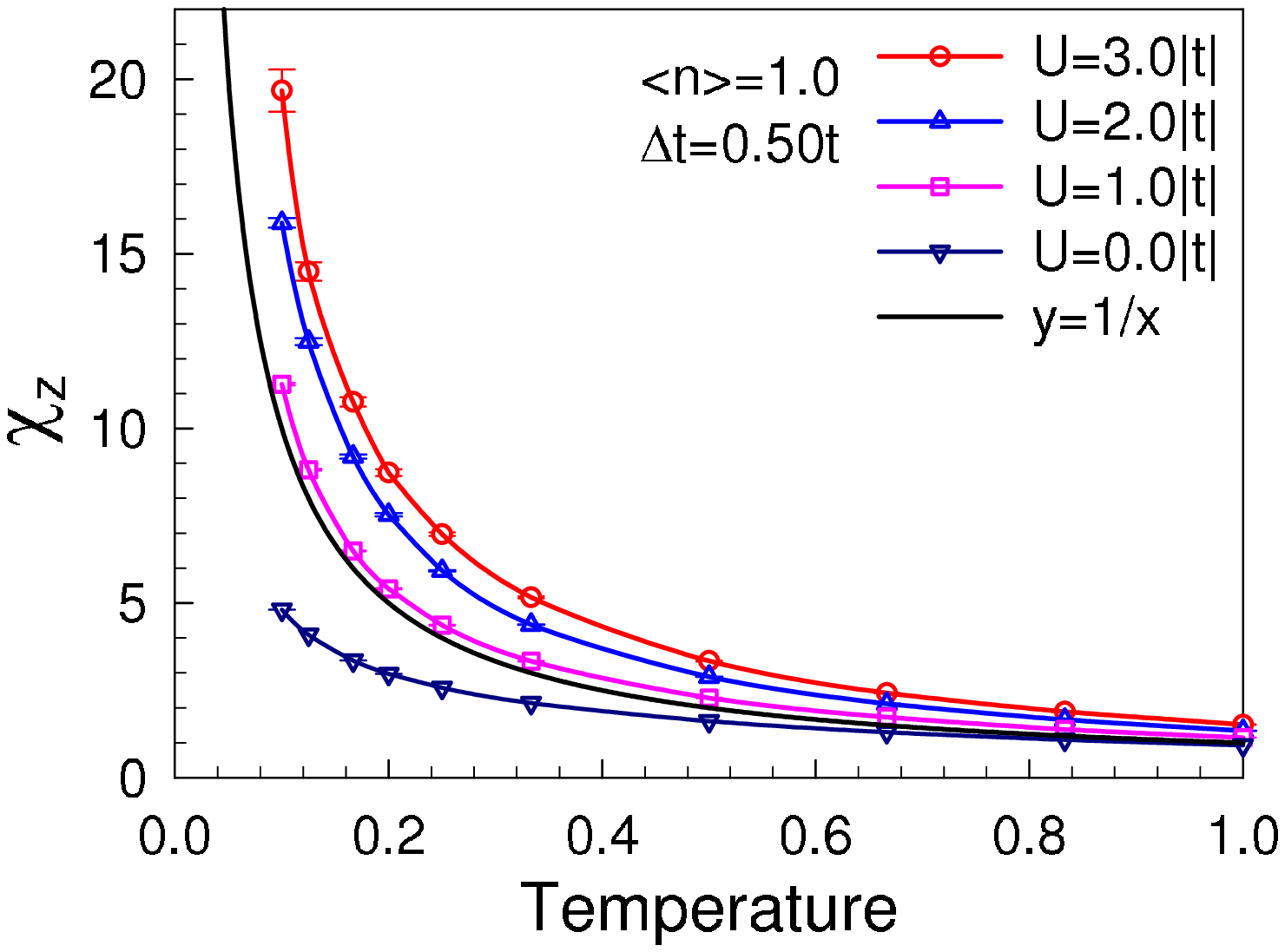}
\caption{(Color online) The $\chi_z$ at $\Delta t=0.50t$ and
$\avg{n}=1.0$ with different $U$, which shows that the
$\chi_z$ is enhanced greatly as the interaction $U$ increases, and
as $U\geq 1.0|t|$, a possible ferromagnetic-like behavior is
predicted where the $\chi_z$ tends to diverge at a relative low
temperature.} \label{differentU}
\end{figure}

For shedding light on the importance of strain, we present the
temperature dependent $\chi_z$ at different strain values in Fig. \ref{strain}.
It is clear seen that the $\chi_z$ is largely enhanced by strain. The strain decreases the value of $t$, and thus enhances
the effective strength of electron-electron interactions $U/t$. As a consequence we
 expect that edge magnetism should be enhanced by strain. This edge-state
 magnetism has already been detected by scanning tunnelling microscopy\cite{Feldner2011}. 

In the calculations we have done, the variation of
hopping parameters depends on the amount of strain, which is a
function of the lattice deformation. The variation of the hopping
parameters dependence on lattice deformation has been studied using
first-principles calculations for a wide range lattice
deformations\cite{Ribeiro2009}.  From the results published in the
literature\cite{Ribeiro2009,Pereira2009}, one may estimate that
$\Delta t=0.3t$ corresponds to deformation $e =  dL/L =15 \%$. Both
$ab-initio$ calculation\cite{Liu2007} and experiments\cite{Kim2009}
show that graphene can sustain reversible deformations of the order
of 20$\%$, which corresponds to  $\Delta t=0.50 t$. For the detail
discussions on the relationship between $\Delta t$ and lattice
deformation, we refer the readers to Refs. [\onlinecite{Ribeiro2009,Pereira2009,Peres2009}].

\begin{figure}[tbp]
\includegraphics[scale=0.45]{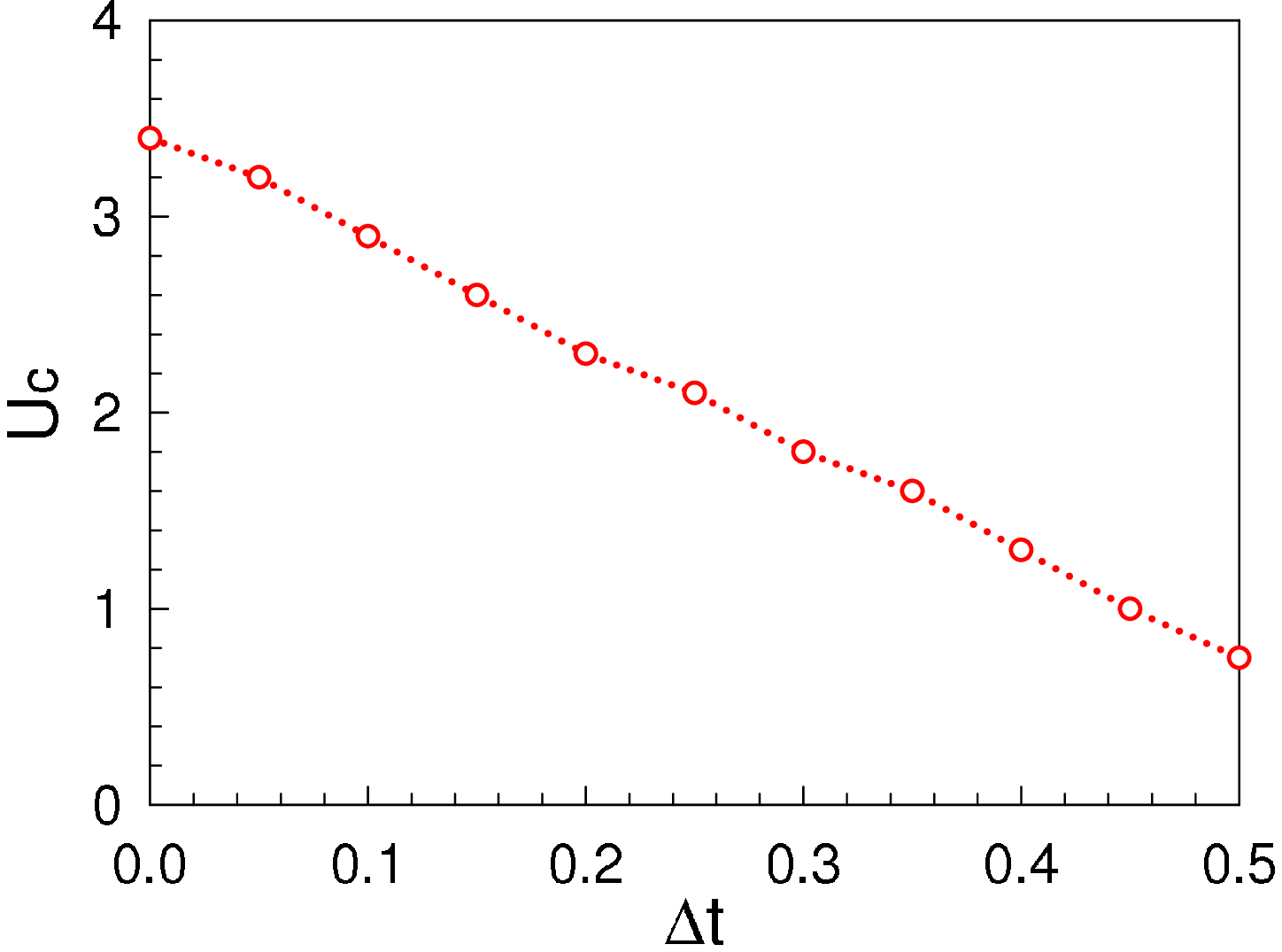}
\caption{(Color online) The critical interaction $Uc$ as a function of strain.}
\label{Uc}
\end{figure}

For understanding the physics induced by the Coulomb
interaction $U$, we compute $\chi_z$ of the graphene quantum dot
with 104 sites for different $U$  values. The results are depicted in Fig. \ref{differentU}.
We can
see that the $\chi_z$ is enhanced by as $U$ increases. At $U=0$,
 $\chi_z$
 behaves like that of a paramagnetic system which does not diverge at a finite low temperature, while as $U>1.0$$\mid$$t$$\mid$, a
ferromagnetic like behavior is shown for $\chi_z$ as $\chi_z$ tends to diverge at a relative low temperature. This indicates that
edge magnetism can be realized in a strained
graphene quantum dot. The physical mechanism that favors
ferromagnetic states at zig-zag edges is as follows: the stress along the
zig-zag edges tends to produce dimmers weakly coupled between them, which
 favors a magnetic state at those tightly bound atoms; this contrasts to what happens
along the armchair edges. On the temperature
 dependent magnetic susceptibility at $U=0$, one can view that the $U=0$ case
 as an extension from the small $U>0$ region. 

In Fig. \ref{Uc}, we plot the critical interaction $U_c$ as a function of
strain. The $U_c$ decreases as the strain increases, and one may estimate an
 {\it optimal} set of parameters as $U=2.3$$\mid$$t$$\mid$ and $\Delta t=0.20t$,
which maybe an ideal value for
the experimental realization.
Let us now discuss the definition of $U_c$. For a very large dot, which is
almost equivalent to the bulk system, the full symmetry of the honeycomb
lattice is restored. In this case a second-order phase transition, at a mean field critical
 Hubbard interaction, can be defined and used to describe the magnetic
transition\cite{Peres2009}. Here, for a finite system, we use the $U_c$ to define the
the crossover where the edge magnetic susceptibility may diverge
at some value of $U$ and strain. For a fixed strain $\Delta t$, we
calculate the temperature dependent magnetic susceptibility at different $U$ values
and extract the temperature $T_c$ where the magnetic susceptibility may diverge.
If the extracted temperature $T_c$ is positive, we define the corresponding lowest
$U$ as $U_c$ for a fixed strain $\Delta t$.

\begin{figure}[tbp]
\includegraphics[scale=0.45]{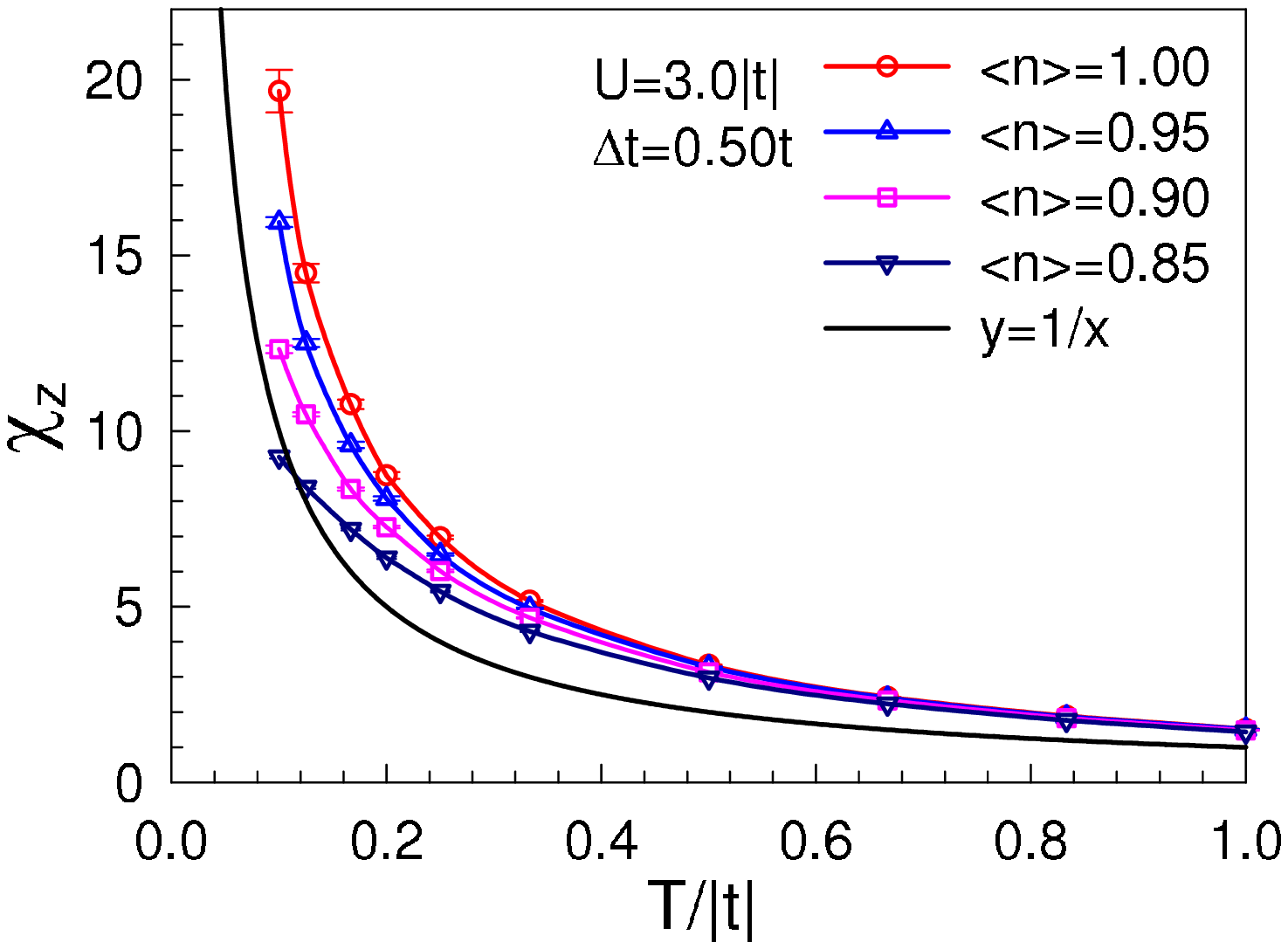}
\caption{(Color online) The $\chi_z$ at $U=3.0$$\mid$$t$$\mid$ and $\Delta t=0.50t$ with
different $\avg{n}$.}
\label{different n}
\end{figure}

In Fig. \ref{different n}, we present $\chi_z$ of a
 graphene quantum dot with 104 sites versus temperature
at different electronic fillings $\avg{n}$.
When the electron filling decreases away from the half filling,
$\chi_z$ decreases slightly at low temperatures, and the
ferromagnetic-like behavior is suppressed when the doping is
larger than 10\%.

\section{Summary of results}
In summary, we have studied the edge sate magnetism of a strained graphene quantum dot
by using the determinant quantum Monte Carlo method. It has been found
found that the magnetic susceptibility $\chi_z$ at the zigzag edge increases
as the temperature decreases. This is specially true in low temperature region.
The susceptibility $\chi_z$ is markedly strengthened  by the on-site Coulomb interaction
and is enhanced  by  strain, which shows a ferromagnetic-like
behavior for a relative small Hubbard interaction $U$  with judicious  choice of strain.
The resultant strongly-enhanced ferromagnetic
fluctuations in graphene quantum dots may facilitate the development of many spintronics
applications.

\section*{acknowledgments}
T. Ma thanks CAEP for partial financial support. This work is supported by NSFCs (Grant. Nos. 11374034 and 11334012), the Fundamental Research Funds for the Central Universities, and is partially supported by the FEDER COMPETE Program and by the Portuguese Foundation for Science
and Technology (FCT) through grant PEst-C/FIS/UI0607/2013. We acknowledge
 support from the EC under Graphene Flagship (contract no. CNECT-ICT-604391).

\end{document}